\documentclass[%
reprint,
 amsmath,amssymb,
 aps,
 prl,
]{revtex4-1}

\usepackage{graphicx}
\usepackage{dcolumn}
\usepackage{bm}
\usepackage{units}

\begin{document}

\preprint{APS/123-QED}

\title{Excitation and Control of Plasma Wakefields by Multiple Laser Pulses}
\author{J. Cowley$^1$}
\author{C. Thornton$^1$}
\author{C. Arran$^1$}
\author{R. J. Shalloo$^1$}
\author{L. Corner$^1$}
\author{G. Cheung$^1$}
\author{C.D. Gregory$^2$}
\author{S.P.D. Mangles$^3$}
\author{N. H. Matlis$^4$}
\author{D. R. Symes$^2$}
\author{R. Walczak$^1$}
\author{S. M. Hooker$^1$}
\email{simon.hooker@physics.ox.ac.uk}

\affiliation{$^1$John Adams Institute for Accelerator Science, University of Oxford, Denys Wilkinson Building, Keble Road, Oxford OX1 3RH, United Kingdom}
\affiliation{$^2$Central Laser Facility, Rutherford Appleton Laboratory, Didcot OX11 0QX, United Kingdom}
\affiliation{$^2$John Adams Institute for Accelerator Science, Blackett Laboratory, Imperial College London, London SW7 2AZ, United Kingdom}
\affiliation{$^4$Deutsches Elektronen-Synchrotron (DESY), Notkestra\ss e 85, Hamburg 22607, Germany}

\date{14 August 2017}

\begin{abstract}
We demonstrate experimentally the resonant excitation of plasma waves by trains of laser pulses. We also take an important first step to achieving an energy recovery plasma accelerator by showing that unused wakefield energy can be removed by an out-of-resonance trailing laser pulse. The measured laser wakefields are found to be in excellent agreement with analytical and numerical models of wakefield excitation in the linear regime. Our results indicate a promising direction for achieving highly controlled, GeV-scale laser-plasma accelerators operating at multi-kilohertz repetition rates.

This article was published in \text{Physical Review Letters} \textbf{119}, 044802 on 27 July 2017. \href{https://doi.org/10.1103/PhysRevLett.119.044802}{DOI: 10.1103/PhysRevLett.119.044802}

\noindent \copyright 2017 American Physical Society.
\end{abstract}

\pacs{Valid PACS appear here}
\maketitle

Particle accelerators lie at the heart of many areas of science, technology, and medicine either through direct application of the particle beams or by driving radiation sources such as synchrotrons and free-electron lasers. With conventional radio-frequency technology the electric field used to accelerate particles is typically less than $\unit[30]{MV m^{-1}}$, which is a significant factor determining the size and cost of the machine. In distinct contrast, plasma accelerators can generate gradients of order $\unit[100]{GV m^{-1}}$, which shrinks the length of the acceleration stage by orders of magnitude.

In a plasma accelerator the acceleration field is generated within a trailing plasma wakefield excited by displacement of the plasma electrons by a driving laser pulse \cite{Tajima:1979,  Malka:2008, Esarey:2009, Norreys:2009hi} or particle bunch \cite{Chen:1985jm, Muggli:2009ip}. Laser-driven plasma accelerators have made impressive progress \cite{Hooker:2013jk} in recent years. They can now generate electron beams with energies comparable to those used in synchrotrons and FELs (a few GeV), but in accelerator stages only a few centimetres long \cite{Leemans:2006, Kneip:2009, Wang:2013el}, with bunch durations in the femtosecond range \cite{Buck:2011dg, Lundh:2011b, Heigoldt:2015cd}, and with properties ideal for generating femtosecond duration visible to X-ray pulses  \cite{Schlenvoigt:2008, Fuchs:2009, Kneip:2010,Cipiccia:2011, Phuoc:2012vb, Powers:2013bx, Khrennikov:2015gx}.

In almost all recent work the plasma wakefield has been driven by single laser pulses from high-power Ti:sapphire chirped-pulse-amplification (CPA) laser systems. Unfortunately, these have very low wall-plug efficiency ($< 0.1\%$) and cannot readily operate at pulse repetition frequencies much above $f_\mathrm{rep} = \unit[1]{Hz}$. At present, therefore, the driver parameters severely restrict the number of potential applications of laser-plasma accelerators.

We recently re-examined \cite{Hooker:2014ij} multi-pulse laser wakefield acceleration (MP-LWFA) in which the wakefield is excited by a train of low energy laser pulses, rather than by a single, high-energy pulse. If the pulses are spaced by the plasma wavelength $\lambda_\mathrm{p0} = 2 \pi c / \omega_\mathrm{p0}$, then the wakefields driven by the pulses in the train will add coherently, causing the plasma wave amplitude to grow towards the back of the train. Here, the plasma frequency is  $\omega_\mathrm{p0} = 2 \pi / T_\mathrm{p0} = (n_\mathrm{e0} e^2 / m_\mathrm{e} \epsilon_0)^{1/2}$, where $n_\mathrm{e0}$ is the ambient electron density.

Using a train of low-energy laser pulses opens plasma accelerators to novel laser technologies, such as fibre or thin-disk lasers, which cannot directly deliver joule-level pulses, but which can provide lower-energy pulses with $f_\mathrm{rep}$ in the kilohertz range, whilst achieving wall-plug efficiencies at least two orders of magnitude higher than conventional solid-state lasers \cite{Klenke:2014kga}. Our recent analysis \cite{Hooker:2014ij} showed that a MP-LWFA driven by a near-term laser system of this type could generate GeV-scale electron bunches at $f_\mathrm{rep} = \unit[10]{kHz}$, and that these could drive compact coherent and incoherent X-ray sources with \emph{average} brightnesses exceeding those available from large scale, non-superconducting, RF accelerators. A further advantage of MP-LWFA is that it provides a natural architecture for ``energy recovery'': the use of one or more trailing laser pulses to remove (and potentially recycle) energy remaining in the wakefield after particle acceleration. Energy recovery is likely to be an important capability in future plasma accelerators operating at high average powers.

In this Letter we present the first demonstration of MP-LWFA in this regime. We also take an important first step towards achieving energy recovery by showing that a suitably delayed laser pulse can damp the plasma wave driven by a leading pulse. We achieve this through measurements of plasma waves by frequency domain holography (FDH) and a new analysis method, Temporally-Encoded Spectral Shifting (TESS) \cite{Matlis:2016ij}; we demonstrate that these two analyses are in excellent agreement, and that our results are well described by a linear response model of wakefield excitation.

\begin{figure}[tb]
\centering
\includegraphics[width = 86mm]{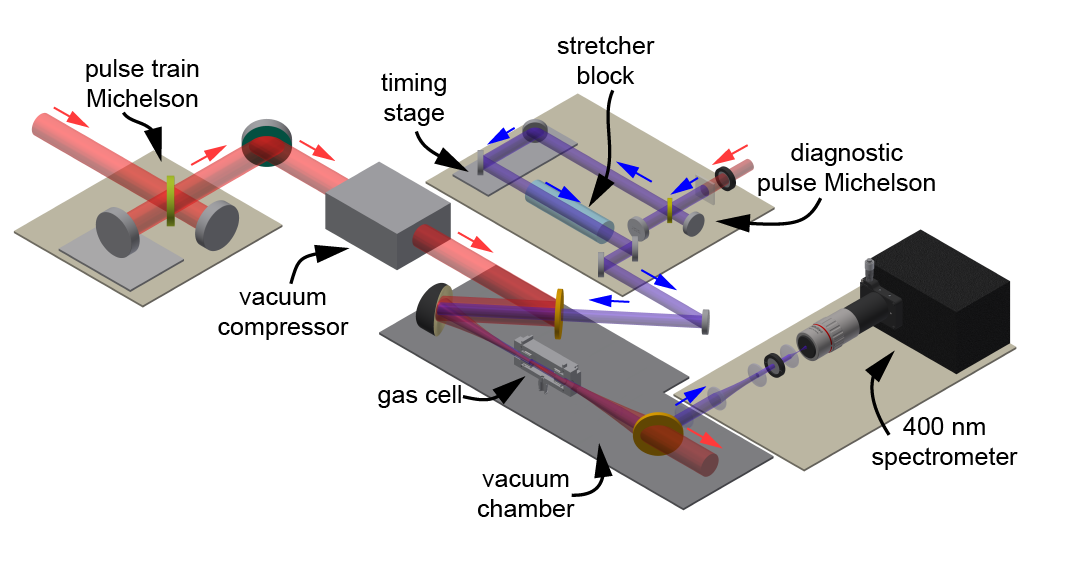}
\caption{Schematic diagram of the experiment layout. The propagation path of the driving pulse train is shown in red, and that of the probe and reference beams is shown in blue. The laser compressor and the components shown above the darker  base are located in the vacuum chamber; all other components are mounted in air.}
\label{Fig:Expt_layout}
\end{figure}

Since laser systems generating directly the pulse trains required for MP-LWFA are still under development, this first demonstration employed a Ti:sapphire laser --- the Gemini (Astra TA2) laser at the Rutherford Appleton Laboratory --- reconfigured to generate trains of laser pulses. In its standard configuration this laser delivers to target approximately \unit[600]{mJ}, \unit[40]{fs} laser pulses with a centre wavelength $\lambda_0 = \unit[800]{nm}$ at $f_\mathrm{rep} = \unit[5]{Hz}$.

Figure \ref{Fig:Expt_layout} shows schematically the experimental arrangement employed (see Supplemental Material at [URL will be inserted by publisher] for further details of the experimental arrangement and analysis methods). Single, temporally-chirped pulses from the laser system were converted into pulse trains by placing a Michelson interferometer between the final laser amplifier and its vacuum compressor \cite{Weling:1996bj}. The pulse train Michelson and compressor combination could be operated in two ways. If the compressor was set to give full compression then their output comprised a pair of short (approximately \unit[50]{fs}) pulses temporally separated by $\delta \tau = \Delta x/c$, where $\Delta x$ is the path difference between the Michelson arms. However, with the compressor set for \emph{partial} compression its output comprised two chirped pulses separated in time by $\Delta x/c$; beating between these created a train of pulses of spacing $\delta \tau = 2 \pi c/ \phi_\mathrm{dr}^{(2)} \Delta x$, where $\phi_\mathrm{dr}^{(2)}$ is the second derivative of the temporal phase of the incident drive pulse. The temporal intensity profiles of the pulse trains were determined by combining a model of the laser compressor and pulse train Michelson with measurements of the pulse train spectrum and single-shot autocorrelation (SSA) \cite{Shalloo:2016fq}.

The pulse train leaving the compressor was directed to an $f = \unit[1]{m}$ off-axis paraboloid, used at f/18, which focused the pulses into a gas cell containing pure hydrogen gas. The spot size ($1/e^2$ radius of the transverse intensity profile) of the focused pulse trains was measured to be $w_0 = \unit[(35 \pm 5)]{\mu m}$.\

\begin{figure}
\centering	
\includegraphics[width = 86mm]{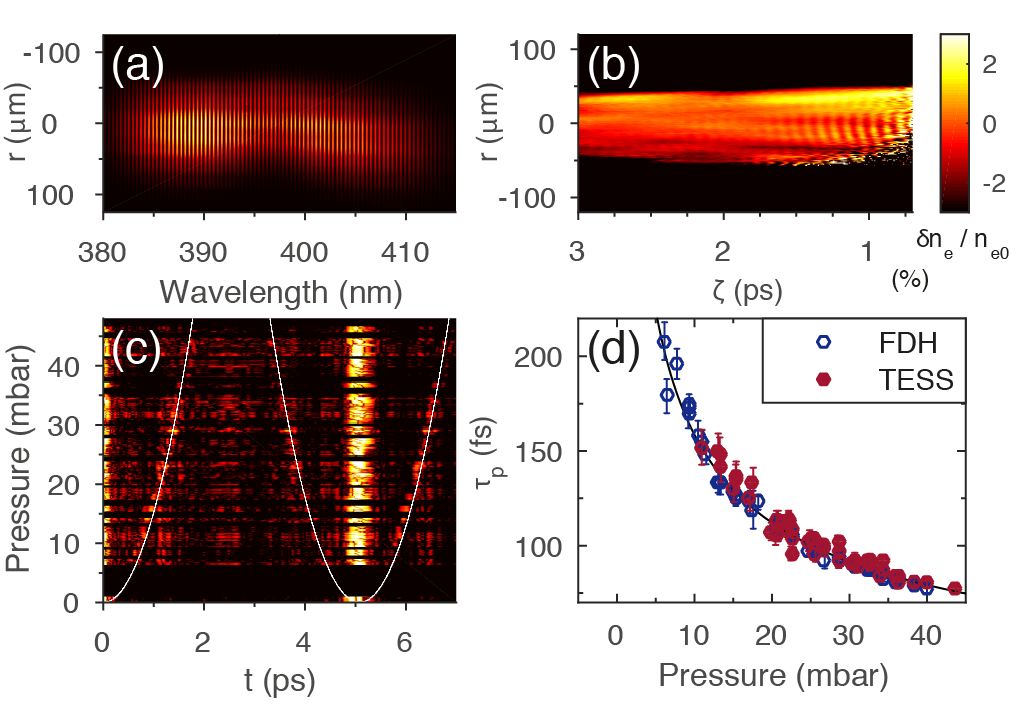}
\caption{FDH and TESS analyses of linear plasma wakefields driven by a single laser pulse of energy approximately \unit[270]{mJ} and pulse duration $\unit[(43 \pm 5)]{fs}$. (a) shows an example spectral interferogram. (b) shows an example of the wakefield recovered by FDH for a cell pressure of $\unit[(31 \pm 1)]{mbar}$, where $\zeta = 0$ corresponds to the centre of the pump pulse. (c) shows a waterfall plot of Fourier transforms of the spectral interferograms, where the magnitude of the Fourier transform is plotted on a logarithmic scale. The dashed white line shows the expected position of the satellites calculated from the expected plasma frequency. (d) shows, as a function of the gas pressure, the plasma period determined by the FDH and TESS analyses. The solid curve is the plasma period calculated assuming an electron density equal to twice the density of hydrogen molecules. The error bars are estimated from the uncertainty in determining the satellite separation in (c) and the plasma period in (b).}
\label{FDH-TESS_comparison}
\end{figure}

Plasma wakefields driven by the pulse train were probed by frequency domain holography \cite{Matlis:2006dv}. In this method a frequency-chirped probe pulse co-propagates with the plasma wave and a reference pulse located ahead of the plasma wave. These diagnostic pulses are then interfered in a spectrograph to give a spectral interferogram, with spatial information in the non-dispersed direction. When the chirped probe pulse interacts with a plasma wave, each of its frequency components experiences a phase shift which depends on the local wakefield amplitude; after a length $\ell$ of plasma this phase shift can be written as $\phi_\mathrm{p}(\zeta) =  \frac{\omega_0}{c} \ell \left[\eta(\zeta) - \eta_0\right]$, where $\omega_0$ is the angular frequency of the probe pulse, $\zeta = t - \ell / c$ ,  $\eta(\zeta)$ is the refractive index of the plasma, and $\eta_0$ is the refractive index experienced by the reference pulse.   The spectrum of the combined transmitted probe and reference pulses comprises spectral fringes of angular frequency separation $\Delta \omega  = 1 / \Delta t$, where $\Delta t$ is the temporal separation of the probe and reference pulses, modulated by a spectral phase $\Delta \psi(\omega)$ which depends on the wakefield, as shown in Fig.\ \ref{FDH-TESS_comparison}(a). Frequency domain holography uses well known Fourier techniques to extract $\Delta \psi(\omega)$ from the interferogram, and hence the temporal phase shift caused by the plasma wave \cite{Matlis:2006dv}.

In this work we also used a TESS analysis of the same data \cite{Matlis:2016ij}, which is applicable when the plasma wave is sinusoidal. In this approach a Fourier transform of the interferogram yields a sideband at $t = \Delta t$ and a series of satellites at $t = \Delta t \pm m \psi^{(2)} \omega_{p0}$  where $m = \pm 1, \pm 2, \pm 3, \ldots$ and $\psi^{(2)}$ is the group delay dispersion (GDD) of the probe and reference pulses. The ratio of the amplitudes of the satellites to the sideband can be shown to be \cite{Arran:2017},

\begin{equation}
	r_m = \frac{J_m(\Delta \phi_\mathrm{p})}{J_0(\Delta \phi_\mathrm{p})} \frac{\mathcal{F}(m \omega_\mathrm{p0})}{{\mathcal{F}(0)}}, \label{Eqn:satellite_ratio}
\end{equation}
where $\Delta \phi_\mathrm{p} = (\omega_\mathrm{p0}^2 / 2 \omega)(\ell /c) (\delta n_\mathrm{e0} / n_\mathrm{e0})$ is proportional to the wake amplitude and,

\begin{equation}
	\mathcal{F}(m \omega_\mathrm{p0}) = \int_{0}^{\infty}\sqrt{S_\mathrm{pr,inc}(\omega + m \omega_\mathrm{p0})} \sqrt{S_\mathrm{ref,inc}(\omega) } \mathrm{d} \omega,\label{Eqn:Spectral_overlap}
\end{equation}
in which $S_\mathrm{pr,inc}(\omega)$ and $S_\mathrm{ref,inc}(\omega)$ are the spectra of the incident probe and reference pulses.

A pair of $\lambda = \unit[400]{nm}$ diagnostic pulses, with an adjustable temporal separation $\Delta t$, were generated by passing a separately-compressed and frequency-doubled fraction of the main laser pulse through a Michelson interferometer. These pulses were chirped and stretched to a duration of around \unit[1.5]{ps} by sending them through a \unit[160]{mm} long block of BK7 glass. The diagnostic pulses were propagated co-linearly with the driving pulse train by directing them through a dichroic mirror; after propagating through the gas cell they were separated from the pulse train by a second dichroic mirror and imaged onto the entrance slit of a spectrograph.

Figure \ref{FDH-TESS_comparison} shows the results of FDH and TESS measurements of the wakes driven by a single laser pulse. An example wakefield retrieved by FDH is shown in Fig.\ \ref{FDH-TESS_comparison}(b): the wake can be observed clearly, with a transverse extent which is compatible with the focal spot size of the driving laser, and with wavefronts which are only slightly curved, which is consistent with a linear wakefield.   The plasma period, read directly from the plot, is found to be $T_\mathrm{p0} = \unit[(90 \pm 5)]{fs}$, which agrees with the expected value of $T_\mathrm{p0} = \unit[(91 \pm 2)]{fs}$ for this cell pressure.

The wake in Fig.\ \ref{FDH-TESS_comparison}(b) can be observed up to $\zeta \approx \unit[2]{ps}$ after the pump pulse, corresponding to approximately 20 plasma periods. Particle-in-cell simulations show that ion motion does not cause a decrease in the wake amplitude until approximately 80 plasma periods after the driving pulse \cite{Hooker:2014ij}. The observed decrease is therefore likely to be caused by variations of the plasma density within the gas cell since, in the presence of such variations, the number of measurable plasma periods will be approximately $n_\mathrm{e0} / (2 \Delta n_\mathrm{e0})$, where $\Delta n_\mathrm{e0}$ is the range of density, and hence the data is consistent with $\Delta n_\mathrm{e0} / n_\mathrm{e0} \approx 2.5\%$.

Figure \ref{FDH-TESS_comparison}(c) shows, as a function of the cell pressure,  a waterfall plot of  Fourier transforms of the spectral interferograms. The sideband at $t = \Delta t \approx \unit[5.1]{ps}$, corresponding to the probe-reference separation, can be seen clearly, as can the $ m = \pm 1$ TESS satellites; the separation of these satellites  --- and also of a satellite to the DC peak at $t = 0$ --- follows closely that expected from the measured GDD of the probe pulse and the plasma frequency calculated from the initial gas pressure, assuming full ionization by the driving laser pulse. The plasma periods determined from the FDH and TESS analyses are compared in Fig.\ \ref{FDH-TESS_comparison}(d) and are seen to be in excellent agreement with each other and with the calculated plasma period.

\begin{figure}[tb]
\centering
\includegraphics[width = 86mm]{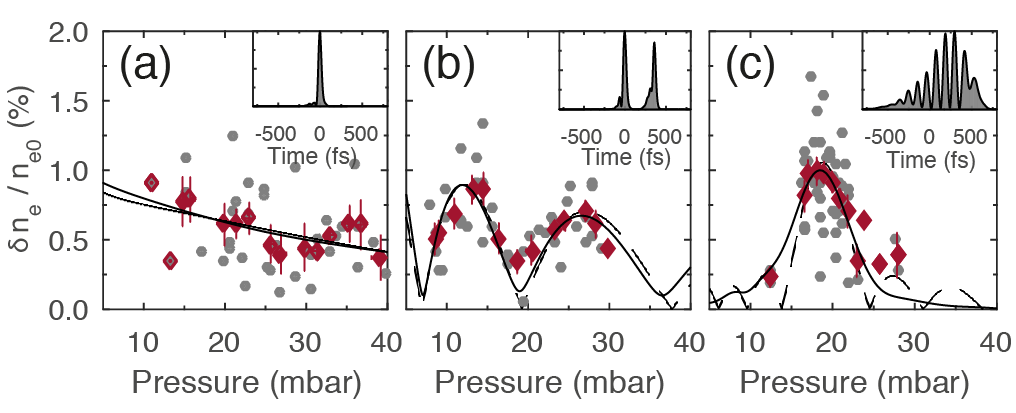}
\caption{Relative wakefield amplitudes, as a function of gas cell pressure, measured at delay $\zeta$ between the centre of the pulse train and the centre of the probe pulse for a driving pulse train comprising $N$ pulses of measured pulse separation $\delta \tau$ and total energy $E$ where: (a)  $N=1$, $E = \unit[270]{mJ}$, $\zeta = \unit[2.2]{ps}$; (b) $N=2$, $\delta \tau = \unit[(365 \pm 40)]{fs}$, $E = \unit[160]{mJ}$, $\zeta = \unit[2.5]{ps}$; and (c) $N \approx 7$, $\delta \tau = \unit[(112 \pm 6)]{fs}$, $E = \unit[170]{mJ}$, $\zeta = \unit[1.3]{ps}$. Gray circles show single measurements and black diamonds show the same data averaged over pressure bins of width \unit[4]{mbar} (a, b) or \unit[2]{mbar} (c); the error bars are standard errors and the $y$-axes are the same for all plots. The insets show the measured driving pulse trains. The dashed lines show fits of eqn \eqref{Eqn:Resonance_multi-pulse}, and the solid lines show the wake amplitudes calculated for the pulse trains shown in the figure insets.}
\label{Fig:Wake_amplitudes}
\end{figure}

Figure \ref{Fig:Wake_amplitudes} shows, as a function of cell pressure, the relative amplitude of the plasma waves driven by trains of  $N = 1$, $N = 2$ and $N \approx 7$ pulses, as determined by TESS analyses. In the linear regime the relative amplitude of the plasma wave driven by a single driving pulse with Gaussian transverse and temporal profiles is \cite{Dorchies:1999vb},

\begin{equation}
\frac{\delta n_\mathrm{e} }{n_\mathrm{e0}} = A \omega_\mathrm{p0} \tau_0  \left[1 + \left(\frac{2 \sqrt{2} c}{\omega_\mathrm{p0} w_0} \right)^2 \right] \exp \left[- \frac{\left(\omega_\mathrm{p0} \tau_0\right)^2}{16 \ln 2}  \right],\label{Eqn:Resonance_single-pulse}
\end{equation}
where  $\tau_0$ is the full-width at half maximum of the temporal profile, and the parameter $A$ is proportional to the peak laser intensity. Figure \ref{Fig:Wake_amplitudes}(a) shows a fit of equation \eqref{Eqn:Resonance_single-pulse} to the data, where $A$ and $\tau_0$ are taken as free parameters and $\omega_\mathrm{p0}$ is calculated from the gas pressure. The fit yields $\tau_0 = \unit[(49 \pm 8)]{fs}$, which is consistent with the value of $\tau_0 = \unit[(46 \pm 7)]{fs}$ measured with the SSA. Fig.\ \ref{Fig:Wake_amplitudes}(a) also shows excellent agreement between the data and a fit to the wakefield amplitude calculated for the measured temporal intensity profile of the driving pulse, the only fitting parameter being the parameter $A$.

From elementary considerations,  in the linear regime the relative amplitude of the wakefield behind a train of $N$ identical driving pulses spaced in time by $\delta \tau$ is,

\begin{equation}
\left(\frac{\delta n_\mathrm{e}}{n_\mathrm{e0}} \right)_N = \left(\frac{\delta n_\mathrm{e}}{n_\mathrm{e0}} \right)_1 \left|\frac{\sin \left( \frac{1}{2} N \omega_\mathrm{p0} \delta \tau\right)}{\sin \left( \frac{1}{2} \omega_\mathrm{p0} \delta \tau\right)}\right|. \label{Eqn:Resonance_multi-pulse}
\end{equation}
Figure \ref{Fig:Wake_amplitudes}(b) shows the measured wake amplitude, as a function of pressure, for a pair of laser pulses. Very clear constructive and destructive interference of the two wakefields is observed, as expected. A fit to  eqn \eqref{Eqn:Resonance_multi-pulse} yields  $\delta \tau = \unit[(289 \pm 5)]{fs}$, which is close to the measured value. Better agreement with the data is obtained if the pressure variation of the wake amplitude is calculated from the measured temporal intensity profile of the driving pulses. For this fit the free parameters were an overall scaling factor for the wake amplitude, and a scaling factor $\alpha$ for the temporal axis of the measured driving pulses, such that $\zeta \rightarrow \alpha \zeta$; the fit yields $\alpha = 0.81$. An analysis of these data shows that the second (smaller) laser pulse reduced the amplitude of the wakefield by approximately $40\%$ (from a relative amplitude of $\delta n_e / n_{e0} = 0.60\%$ to $0.35\%$); this energy will be removed from the plasma in the form of blue-shifted photons in the trailing laser pulse \cite{Murphy:2006, Dias:1998qi}.

Figure \ref{Fig:Wake_amplitudes}(c) shows the measured wake amplitude as a function of the cell pressure for  $N \approx 7$ laser pulses. A pronounced resonance is observed when the plasma period matches the pulse spacing $\delta \tau$. Also shown is a fit of equation \eqref{Eqn:Resonance_multi-pulse} for a train of $N = 7$ identical pulses. Once again excellent agreement between the data and analytical theory is obtained, the fit yielding  $\delta \tau = \unit[(116 \pm 2)]{fs}$ which agrees with the measured value. The solid line shows the variation of the wake amplitude calculated for the measured pulse train, the fit yielding $\alpha = 1.04$.  It is noticeable that the pressure variation of the wake amplitude calculated for the measured pulse train does not exhibit subsidiary maxima; this difference is caused by the small variation of the pulse spacing, and the presence of temporal wings, in the measured pulse train.

We now place this work in context with earlier studies. The MP-LWFA approach is closely related to the plasma beat-wave accelerator (PBWA) \cite{Tajima:1979, Joshi:1984vf},  in which two long laser pulses of angular frequencies $\omega_1$ and $\omega_2 =\omega_1 + \omega_\mathrm{p0}$ are combined to form a pulse modulated at $\omega_\mathrm{p0}$. Beat-wave excitation of plasma waves \cite{Dangor:1990fx, Clayton:1985, Amiranoff:1992}, and their application to accelerating electrons \cite{Clayton:1993, Tochitsky:2004}, have both been demonstrated. To counter the saturation of the wakefield amplitude caused by the relativistic increase in electron mass \cite{Rosenbluth:1972us},  Deutsch et al.\ proposed \cite{Deutsch:1991} frequency chirping one or both of the laser pulses to maintain resonance as the wake grows in amplitude.

The MP-LWFA concept  has been investigated theoretically \cite{Nakajima:1992, Berezhiani:1992vw, Umstadter:1994, Johnson:1994uo, Dalla:1994uta, Bonnaud:1994uo, Cairns:1995wa, Umstadter:1995dk, Kalinnikova:2008} but has not previously been demonstrated. It can be considered to be a generalization of the beat-wave scheme since, in principle: the pulses in the train do not have to be mutually coherent; the durations of the pulses are not necessarily related to their separation; and the separations of the pulses within the pulse train can be adjusted to avoid saturation of the plasma wave.

The $N \approx 7$ pulse train shown in Fig.\ \ref{Fig:Wake_amplitudes}(c) is an example of a beat-wave, since it was generated by superposition of two laser pulses of different local frequency, whereas  the $N = 2$ train shown in Fig.\ \ref{Fig:Wake_amplitudes}(b) is more naturally described as MP-LWFA. As far as we are aware, the results presented here for $N \approx 7$ pulses are the first demonstration of beat-wave excitation of a plasma wave with chirped laser pulses. Since in this experiment the wakefields were linear, it was only necessary to ensure that the chirps of the two pulses were approximately equal so that the variation of the beat frequency during the pulse train was small. However, it would be possible to drive larger amplitude wakefields by controlling the chirp \cite{Verluise:2000uz} of one or both pulses so as to maintain resonance with the plasma wave as its amplitude grows.

In summary, we have demonstrated that plasma wakefields can be driven by trains of laser pulses, and that their amplitudes can be controlled by adjusting the laser pulse spacing relative to the plasma wavelength. In addition we have shown that unused wakefield energy can be removed by a trailing laser pulse, which is an important first step towards energy recovery. Our results indicate a route to achieving highly controlled, GeV-scale laser-plasma accelerators operating at multi-kilohertz repetition rates and driven by novel, efficient laser technologies \cite{Hooker:2014ij}. In addition to stimulating new work on the development of laser-plasma accelerators, these results will be of interest to those working on driving plasma accelerators driven by trains of \emph{particle} bunches \cite{Muggli:2008jd, Kallos:2008ib} or self-modulated proton beams \cite{Caldwell:2009, Caldwell:2011dz}.

This work was supported by the UK Science and Technology Facilities Council (STFC UK) [grant number ST/J002011/1]; the Engineering and Physical Sciences Research Council [Studentship No.\ 702499]; the Helmholtz Association of German Research Centres [Grant number VH- VI-503]; and Air Force Office of Scientific Research, Air Force Material Command, USAF [Grant number FA8655-13-1-2141].


%

\end{document}